\documentclass[aps,prl,twocolumn,showpacs]{revtex4}

\usepackage{graphicx}

\begin{document}

\title{
Reversible magnetization and critical fluctuations in
systematically doped YBa$_2$Cu$_3$O$_{7-\delta}$ single crystals}

\author{Hong Gao$^1$, Cong Ren$^1$, Lei Shan$^1$, Yue Wang$^1$, Yingzi Zhang$^1$, Shiping Zhao$^2$, Xin Yao$^3$, Hai-Hu Wen$^1$}

\affiliation{$^1$National Laboratory for Superconductivity,
Institute of Physics and Beijing National Laboratory for Condensed
Matter Physics, Chinese Academy of Sciences, P.~O.~Box 603,
Beijing 100080, P.~R.~China}

\affiliation{$^2$Laboratory for Extreme Conditions, Institute of
Physics and Beijing National Laboratory for Condensed Matter
Physics, Chinese Academy of Sciences, P.~O.~Box 603, Beijing 100080,
P.~R.~China}

\affiliation{$^3$Department of Physics, Shanghai Jiaotong
University, Shanghai, P.~R.~China}

\date{\today}

\begin{abstract}
The temperature and field dependence of reversible magnetization
have been measured on a YBa$_2$Cu$_3$O$_{7-\delta}$ single crystal
at six different doping concentrations. It is found that the data
above 2 T can be described by the scaling law based on the GL-LLL
(lowest Landau level approach based on Ginzburg-Landau theory)
critical fluctuation theory yielding the values of the slope of
upper critical field $-\mathrm{d}H_{\mathrm{c2}}(T)/\mathrm{d}T$
near $T_\mathrm{c}$. This set of values is self-consistent with that
obtained in doing the universal scaling for the six samples. Based
on a simple Ginzburg-Landau approach, we determined the doping
dependence of the coherence length $\xi$ which behaves in a similar
way as that determined from $\xi= \hbar v_\mathrm{F}/E_\mathrm{sc}$
with $E_\mathrm{sc}$ the superconducting energy scale. Our results
may suggest a growing coherence length towards more underdoping.
\end{abstract}

\pacs{74.40.+k, 74.25.Ha, 74.72.Bk, 74.25.Op}

\maketitle

In hole doped high temperature superconductors the transition
temperature $T_\mathrm{c}$ and the maximum quasiparticle gap (or
called as the pseudogap) behave in an opposite way: the former drops
down but the latter rises up towards more underdoping\cite{Timusk}.
Although consensus has been reached on the doping dependence of some
quantities, such as the transition temperature $T_\mathrm{c}$, the
superfluid density $\rho_\mathrm{s}$ and the condensation energy,
etc., it remains still highly controversial about the doping
dependence of the upper critical field or the coherence length in
the underdoped region. In practice, however, to directly determine
$H_\mathrm{c2}(0)$ has turned out to be a difficult task due to its
very large values. An alternatively way to derive
$-\mathrm{d}H_{\mathrm{c2}}(T)/\mathrm{d}T$ near $T_\mathrm{c}$ is
to measure the reversible magnetization or conductivity and then
analyze the data based on the critical fluctuation theory. Using the
Lawrence-Doniach model for layered structure of superconductors,
Ullah-Dorsey obtained expressions for the scaling functions of
various thermodynamic and transport quantities around $T_\mathrm{c}$
\cite{UllahPRL90}. Moreover, Te\v{s}anovi\'{c} et al. pointed out
that the scaling of magnetization due to critical fluctuations near
$H_\mathrm{c2}(T)$ can be represented in terms of the
Ginzburg-Landau (GL) mean field theory on a degenerate manifold
spanned by the lowest Landau level (LLL) \cite{XingPRL92}.  By using
a nonperturbative approach to the Ginzburg-Landau free energy
functional, $M(T)$ curves are evaluated explicitly for quasi-2D
superconductors in a close form as:

\begin{equation}
\frac{M}{(HT)^{1/2}}=Bf\left[{A\frac{T-T_\mathrm{c}(H)}{(HT)^{1/2}}}\right],
\end{equation}
\begin{equation}
f(x)=x-\sqrt{x^2+2},
\end{equation}
where $A$ and $B$ are independent of $T$ and $H$, but A is dependent
on both the GL parameter $\kappa$ and
$|\mathrm{d}H_{\mathrm{c2}}/\mathrm{d}T|_{T_\mathrm{c}}$, $B$
depends on $\kappa$. This scaling behavior is expected specially in
a high magnetic field. Many experiments were tried to test these
scaling laws and obtain the values of the mean-field transition
temperature $T_\mathrm{c}(H)$ and the slope
$-\mathrm{d}H_{\mathrm{c2}}/\mathrm{d}T$
\cite{WelpPRL91,LiPRB,WenPRL99}. However, due to the sample
diversity, the scaling produced values of $T_\mathrm{c}(H)$ and
$-\mathrm{d}H_{\mathrm{c2}}/\mathrm{d}T$ that did not agree with
each other. As a consequence, the universal scaling for
superconducting diamagnetization fluctuations is still elusive. In
this paper, we systematically investigate the diamagnetization
fluctuations in the vicinity of $T_\mathrm{c}$ of \emph{a
YBa$_2$Cu$_3$O$_{7-\delta}$ single crystal} with six oxygen doping
levels. The results indicate the plausibility of the existence of a
universal scaling in the framework of 2D GL-LLL approximation
theory. The doping dependence of $H_\mathrm{c2}(0)$ are also
reliably obtained.

The YBa$_2$Cu$_3$O$_{7-\delta}$ single crystal used here was grown
by top-seeded solution-growth using the Ba3-Cu5-O solvent. Details
for crystal growth were presented elsewhere \cite{YaoX}. It has a
shape of platelet with lateral dimensions of 3.30 mm $\times$ 1.98
mm, thickness of 0.44 mm, and a mass around 15.56 mg.  The different
concentrations of oxygen were achieved by post-annealing the sample
at different temperatures in flowing gas followed by a quenching in
liquid nitrogen. The detailed annealing procedures are as follows:
the as-grown YBa$_2$Cu$_3$O$_{7-\delta}$ single crystal was first
annealed at 400 $^\circ $C for 180 hours with flowing oxygen and
slowly cooled down to room temperature.  The resulted crystal (S1)
is close to optimally doped with a $T_\mathrm{c}=92.0$ K. Then the
following doping status on this specific sample were achieved by
annealing it in flowing oxygen in sequence: $T_\mathrm{c}=85.3$ K
(520 $^\circ$C for about 110 h, S2), 79.5 K (540 $^\circ $C for 120
h, S3), 68.5 K (580 $^\circ$C for 110 h, S4), 58.6 K (680 $^\circ $C
for 130 h, S5). The last sample (S6) was annealed at 520 $^\circ $C
for 90 h with flowing N$_2$ gas yielding $T_\mathrm{c}=30.5$ K.

The magnetization was measured by a Quantum Design superconducting
quantum interference device (SQUID) magnetometer in both so-called
zero-field-cooled (ZFC) and field-cooled (FC) modes with fields
ranging from 10 G to 5 T parallel to $c$-axis. In the reversible
regime, the data measured using ZFC and FC modes coincide very well.
In SQUID measurement, a scanning length of 3 cm was taken, the SQUID
response curves in reversible regime were fully symmetric to avoid
artificial signal.

\begin{table}
\begin{center}
Table I. Parameters at six different annealed states\\

\begin{tabular}{ccccc}

\hline \hline
sample\qquad   & $T_\mathrm{c}(K)$   & \quad $p$    &$T_{c0}(K)$   & $\left|\frac{\mathrm{d}H_\mathrm{c2}}{\mathrm{d}T}\right|(T/K)$\\
\hline
S1     & $92.0$       & $\quad0.146$      & $\quad 92.7\pm0.6\quad$        & $3.45\pm0.01$  \\
S2     & $85.3$       & $\quad0.127$      & $\quad85.7\pm0.3\quad$        & $3.23\pm0.01$   \\
S3     & $79.5$       & $\quad0.117$      & $\quad80.0\pm0.5\quad$        & $3.03\pm0.01$   \\
S4     & $68.5$       & $\quad0.103$      & $\quad70.1\pm0.5\quad$        & $2.00\pm0.01$  \\
S5     & $58.6$       & $\quad0.093$      & $\quad62.6\pm0.5\quad$        & $1.82\pm0.02$  \\
S6     & $30.5$       & $\quad0.070$      & $\quad36.7\pm1.0\quad$        & $1.43\pm0.05$  \\

\hline \hline

\end{tabular}
\end{center}\label{tab:tableI}
\end{table}

The $T_\mathrm{c}$s' of the sample at six annealed states were
determined by the deviating point of magnetization from the normal
state background\cite{WelpPRL89} (as shown in Fig.1 (a), ZFC mode,
H=10 Oe). One can see that the superconducting transitions are
rather sharp near the transition point. The gradually rounded foot
of $M(T)$ curves in more underdoped samples may be attributed to the
easy flux motion since the system becomes more 2D like. The
relationship between the $T_\mathrm{c}$s' and the oxygen doping
level $p$ [determined by using the phenomenological relation
$T_\mathrm{c}/T_\mathrm{c}^{\mathrm{max}}=1-82.6(p-0.16)^2$ taking
$T_\mathrm{c}^{\mathrm{max}}=93.6 K$] has been summarized in Table
I.

\begin{figure}
\includegraphics[width=8cm]{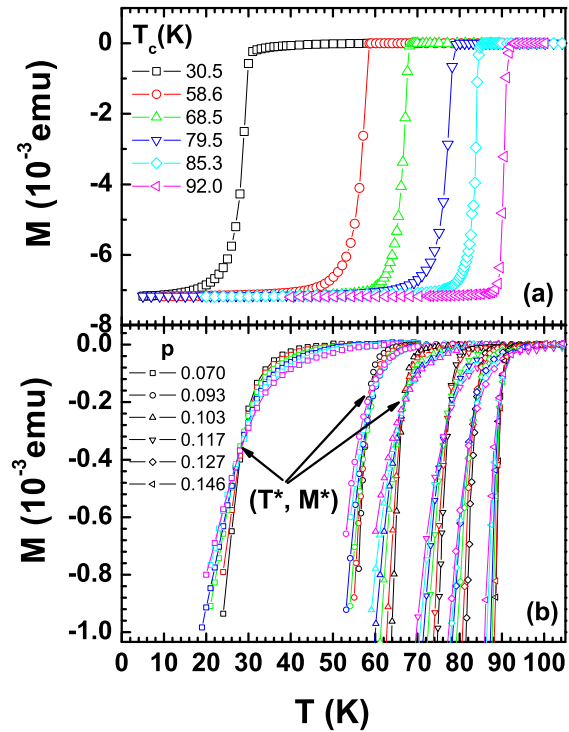}
\caption{(color online)(a) Temperature dependence of the
magnetization measured in the warming-up process with a ZFC mode at
$H=10$ Oe for the sample at six different doping levels. (b)
Magnetization vs. temperature at magnetic fields of 0.5, 1, 2, 3, 4,
5 T. It is easy to see that a crossing point (or a narrow area)
marked as (T*, M*) appears for each sample showing a typical
behavior of critical fluctuation. } \label{fig1}
\end{figure}

Presented in Fig.1(b) are the M(T) curves in the reversible regime
under the applied magnetic fields $H = $0.5, 1, 2, 3, 4, 5 T. A
crossing point at ($T^*$, $M^*$) appears in each set of $M(T)$
curves. Such crossing behavior of the $M(T)$ curves in high magnetic
field has been well described by 2D or 3D GL-LLL scaling theory, and
is a general consequence of fluctuations in the vortex state. It is
interesting to note, however, that the value of $M^*/T^*$ is not a
constant\cite{XingPRL92} in our six sets of $M(T)$ data, being
contradicting with the theoretical prediction\cite{Bulae}. This
deviation from the prediction of the 2D GL-LLL scaling theory were
widely observed in Y123 \cite{a12YBCO}, Hg1223 \cite{Naughton} and
Bi2212 single crystals \cite{a14Bi2212} and may be attributed to the
doping induced change of the anisotropy.

\begin{figure}
\includegraphics[width=8cm]{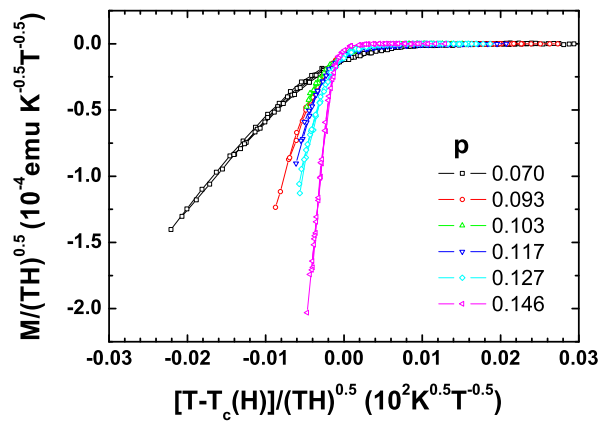}
\caption{(color online)Scaling curves based on the 2D GL-LLL
critical fluctuation theory (see text) for the sample at six
different doping levels. From this scaling one can determine the
slope $|\mathrm{d}H_{\mathrm{c2}}(T)/\mathrm{d}T|_{T_\mathrm{c}}$
for each sample. Interestingly, a new type of crossing point exists
for the scaled curves of different samples.} \label{fig2}
\end{figure}

Despite the deviation mentioned above, excellent 2D scaling curves
are obtained for each set of data with different $p$ and $H\geq 2
$ T, as shown in Fig.~2 where $M(H,T)/(TH)^{1/2}$ is scaled as a
function of the variable of $[T-T_\mathrm{c}(H)]/(TH)^{1/2}$. We
also performed 3D scaling of $M(T)$ for our six samples. The
quality of the 2D scaling are, however, better than that of 3D
scaling for the five sets of $M(T)$ curves of S2, S3, S4, S5 and
S6. For S1 (close to optimally doped) the quality of 3D scaling is
as good as that of 2D scaling with a narrower scaling region. This
may suggest a 2D-3D crossover between $p=0.127$ and 0.146 for our
deoxgenated YBCO crystals.

To fulfill this 2D scaling, two variables $T_{\mathrm{c0}}$ and
$\mathrm{d}H_{\mathrm{c2}}/\mathrm{d}T$ are employed in
$T_\mathrm{c}(H)$ as the fitting parameters:
$T_\mathrm{c}(H)=T_{\mathrm{c0}}-H(\mathrm{d}H_{\mathrm{c2}}/\mathrm{d}T)^{-1}$.
The values of $T_{\mathrm{c0}}$ and
$-\mathrm{d}H_{\mathrm{c2}}/\mathrm{d}T$ resulted from the fit are
also listed in Table I. The critical fluctuation region $\delta
T\equiv T_{\mathrm{c0}}-T_\mathrm{c}$ increases with decreasing $p$,
indicating a larger fluctuation regime for more underdoped YBCO. As
shown in Table I, both the $T_{\mathrm{c0}}$ and
$|\mathrm{d}H_{\mathrm{c2}}/\mathrm{d}T|_{T_\mathrm{c}}$ drops down
towards more underdoping. Another interesting phenomenon shown by
Fig.2 for the scaled curves is that there is a new type of crossing
point for different samples. This new type of crossing point may
indicate a universal scaling among different samples.

In the following we will check the feasibility of the universal
scaling for six samples. As all $M(T)$ data are measured on the
same platelet of YBCO single crystal, this allows us to do
universal scaling for six sets of $M(T)$ curves. The vertical axis
$M/(TH)^{0.5}$ ($y$-axis) is intact because our experiments were
done on the same sample with different doping concentrations and
the GL parameter $\kappa$ is weakly doping dependent. According to
the 2D LLL-scaling theory, an analytical formula of the $x$-axis
for the scaling is written as \cite{XingPRL92}:

\begin{equation}
x=\frac{L}{\kappa}\left|\frac{dH_\mathrm{c2}}{dT}\right|_{T_\mathrm{c}}\left[
\frac{T-T_\mathrm{c}(H)}{(TH)^{1/2}}\right]
\end{equation}
where $L$ is a constant, related to $s$, the interlayer spacing.
From this expression for $x$ axis, it is evident that a \emph{full}
2D scaling analysis of diamagnetization fluctuations has to include
a material-dependent scaling factor $A=\frac{L}{\kappa}\left|
\frac{\mathrm{d}H_{\mathrm{c2}}}{\mathrm{d}T}\right|_{T_\mathrm{c}}$.
Based on the 2D LLL-scaling curve in Fig.~2, such full 2D scaling is
performed by multiplying the ``$x$-axis'' of each 2D LLL-scaled
curve in Fig.~2 by a factor $A'=A(p)/A(p=0.127)$ to make all curves
collapse on a single branch. Here we use the values of
$|\frac{\mathrm{d}H_{\mathrm{c2}}}{\mathrm{d}T}|_{T_\mathrm{c}}$
obtained in 2D scaling analysis and leave $\kappa$ as a free fitting
parameter. In doing the universal scaling we put the 2D-scaled data
of S2 as a reference, i.e., the ``$x$-axis'' is kept as it is. The
resulted collapse of curves from the six samples is shown in Fig. 3.
together with the doping dependence of $\kappa$. It is found that
the scaling quality is predominantly controlled by
$|\mathrm{d}H_{\mathrm{c2}}/\mathrm{d}T|_{T_\mathrm{c}}$ and
$\kappa$ depends weakly on the doping level. Within the experimental
uncertainty the quality of the data collapse is reasonably good. For
a comparison, a theoretical curve generated using Eq.~(2) is also
plotted in Fig.~3. This theoretical curve describes the data rather
well except for a deviation for the sample $p=0.146$
($T_\mathrm{c}=92.0$) on which we believe the system becomes more 3D
like.

\begin{figure}
\includegraphics[width=8cm]{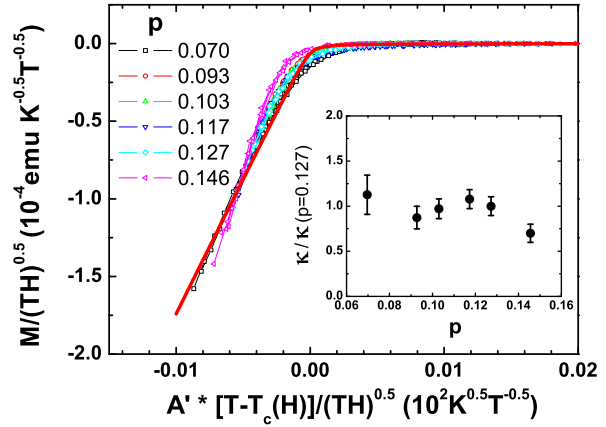}
\caption {(color online)A universal scaling curve by collapsing
the six scaling curves for different doping concentrations based
on the closed-form of the scaling equation of Te\v{s}anovi\'{c} et
al.\cite{XingPRL92}. The thick solid line is a theoretical curve
after Eq.(2). The inset shows the relative $\kappa$ value derived
from doing the universal scaling.} \label{fig3}
\end{figure}

We now discuss the results obtained by the universal scaling
analysis. According to the Werthamer-Helfand-Hohenberg
(WHH)\cite{WHH} theory for a 2D system, the value of
$H_\mathrm{c2}(0)$ is given by $H_\mathrm{c2}(0)=0.69 T_\mathrm{c}
\times |\mathrm{d}H_{\mathrm{c2}}/\mathrm{d}T|_{T_\mathrm{c}}$.
Displayed in Fig. 4(a) is the dependence of $H_\mathrm{c2}(0)$ on
$p$ based on the obtained values of $|dH_{c2}/dT|_{T_\mathrm{c}}$
and $T_\mathrm{c}$ in Table I. Our data show a rough linear
correlation between $H_\mathrm{c2}(0)$ and $p$:
$H_\mathrm{c2}(0)=2620(p-0.058)$. We noted that such linear
$H_\mathrm{c2}(0)-p$ was also obtained for underdoped Bi-2212
polycrystal \cite{Kim2005}. The spin-ordering quantum transition
theory predicts that $H_\mathrm{c2}(0)$ increases with doping in the
underdoped regime \cite{theory}. Quantitatively, this theory
certainly deserves consideration in interpreting our data.

It is interesting to note that in the underdoped regime the
linearity of $H_\mathrm{c2}(0)-p$ leads to a linear dependence of
superfluid density $\rho_s(0)$ on $p$ by the relation:
$H_\mathrm{c2}(0)=\Phi_0/2\pi \xi^2=\Phi_0\kappa^2/2\pi
\lambda^2_{ab}$. Based on the fact that $\kappa$ is weakly $p$
dependent, which is indeed the case for our samples and also found
previously in underdoped YBCO \cite{Gray92}, it is easy to have
$\lambda^{-2}_{ab}(0)\propto \rho_\mathrm{s}(0)\propto p$. This
linear correlation was recently verified in Bi2212 \cite{Kim2005},
La$_{2-x}$Sr$_x$CuO$_4$ and HgBa$_2$CuO$_{4+\delta}$ \cite{Chu}.

\begin{figure}
\includegraphics[width=8.5cm]{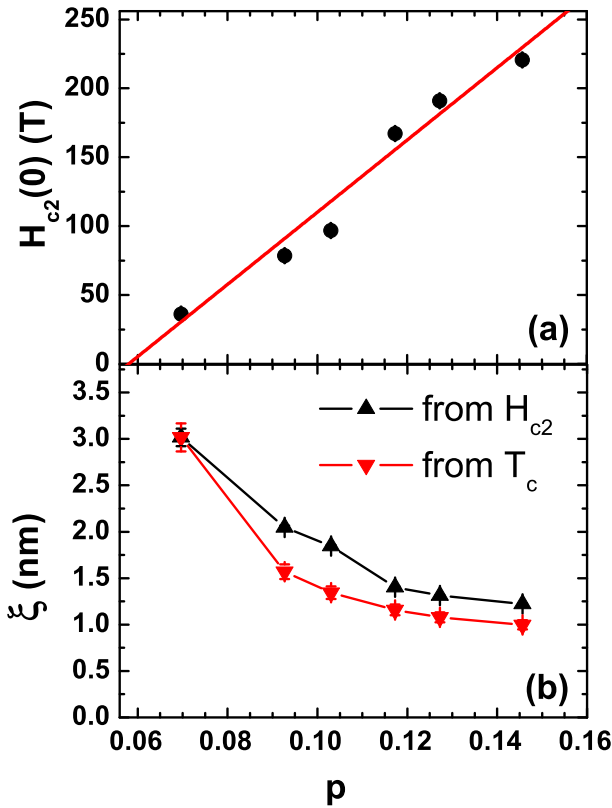}
\caption{(color online)(a)The doping dependence of the upper
critical field $H_\mathrm{c2}(0)$ (phase coherence) derived by
doing the scaling. The solid line is a guide to the eye. (b)Doping
dependence of the coherence length determined from two different
ways (see text).} \label{fig4}
\end{figure}

The coherence length $\xi_{ab}(0)$ of each sample can also be
extracted from the $H_\mathrm{c2}(0)$ value of Fig. 4(a). For
example, we have $\xi_{ab}(0)=[\Phi_0/2\pi
H_\mathrm{c2}(0)]^{1/2}=12.2$ \AA\ for $H_\mathrm{c2}(0)=220$ T for
the sample of $T_\mathrm{c}=92.0$ K. In Fig. 4(b) $\xi_{ab}$ is
summarized and depicted as a function of $p$ in the underdoped
regime.  As $H_\mathrm{c2}(0)$ is reduced to zero at the point
$p=0.058\pm 0.002$, very close to the critical point $p_c=0.050$ for
superconductivity. This implies that $H_\mathrm{c2}(0)$ and
$T_\mathrm{c}$ simultaneously drop to zero at the critical point of
the phase diagram, indicative of the complete suppression of
$\rho_\mathrm{s}(0)$ at $p_\mathrm{c} \approx 0.05$. Another
consistent way to reckon the coherence length is using $\xi=\hbar
v_\mathrm{F}/E_\mathrm{sc}$ with $E_\mathrm{sc}=n
k_\mathrm{B}T_\mathrm{c}$ the superconducting energy
scale\cite{WenPRB2005}, $v_\mathrm{F}$ is the Fermi velocity taking
$2.5\times10^7 cm/s$\cite{Mesot} and is almost doping independent.
In Fig.4(b) we present also the coherence length calculated in this
way with $n \approx 20$. It is clear that the coherence length
obtained in these two ways coincide rather well. Our results here
may indicate a growing coherence length in more underdoped region,
being consistent with our earlier report\cite{WenEPL2003}.

In summary, we have systematically investigated the critical
fluctuations on a YBa$_2$Cu$_3$O$_{7-\delta}$ single crystal at six
different doping concentrations. It is found that the data above $2
T$ can be described by the universal scaling law based on the 2D
GL-LLL critical fluctuation theory. Thus the values of the slope of
upper critical field $-\mathrm{d}H_\mathrm{c2}(T)/\mathrm{d}T$ near
$T_\mathrm{c}$ (and thus $H_\mathrm{c2}(0)$) can be reliably
extracted. The coherence length derived from both $H_\mathrm{c2}(0)$
and $\xi=\hbar v_\mathrm{F}/E_\mathrm{sc}$ show a similar growing
behavior towards more underdoping.

This work is supported by the National Science Foundation of
China, the Ministry of Science and Technology of China (973
project: 2006CB01000), the Knowledge Innovation Project of Chinese
Academy of Sciences.

Corresponding author hhwen@aphy.iphy.ac.cn.

\end{document}